\documentclass[aps,pra,twocolumn,reprint,showpacs,superscriptaddress]{revtex4-1}

\usepackage{amsmath,amssymb,graphicx,color}
\usepackage{times}
\usepackage[table]{xcolor}
\usepackage{amsmath}
\usepackage{graphicx}
\usepackage{amssymb}
\usepackage{txfonts,color,cancel}
\usepackage[colorlinks=true,linkcolor=blue]{hyperref}

\usepackage[normalem]{ulem}

\newcommand\ket[1]{\left|\textstyle{#1}\right\rangle}
\newcommand\bra[1]{\left\langle\textstyle{#1}\right|}
\newcommand\braket[1]{\left\langle\textstyle{#1}\right\rangle}
\newcommand\half{\frac{1}{2}}
\newcommand\down{\downarrow}
\newcommand\up{\uparrow}

\newcommand\Tr{\textrm{Tr}}

\newcommand\Hr{H_\textrm{Rabi}}
\newcommand\Hnp{H_\textrm{np}}
\newcommand\acl{\alpha_\textrm{cl}}
\newcommand\aq{\alpha_\textrm{q}}

\begin{document}

\title{Dissipative Phase Transition in the Open Quantum Rabi Model}

\author{Myung-Joong Hwang}
\affiliation{Insitut f\"ur Theoretische Physik and IQST, Albert-Einstein-Allee 11, Universit\"at Ulm, D-89069 Ulm, Germany}
\author{Peter Rabl}
\affiliation{Vienna Center for Quantum Science and Technology, Atominstitut, TU Wien, 1040 Vienna, Austria}
\author{Martin B. Plenio}
\affiliation{Insitut f\"ur Theoretische Physik and IQST, Albert-Einstein-Allee 11, Universit\"at Ulm, D-89069 Ulm, Germany}

\begin{abstract}
We demonstrate that the open quantum Rabi model (QRM) exhibits a second-order dissipative phase transition (DPT) and propose a method to observe this transition with trapped ions. The interplay between the ultrastrong qubit-oscillator coupling and the oscillator damping brings the system into a steady-state with a diverging number of excitations, in which a DPT is allowed to occur even with a finite number of system components. The universality class of the open QRM, modified from the closed QRM by a Markovian bath, is identified by finding critical exponents and scaling functions using the Keldysh functional integral approach. We propose to realize the open QRM with two trapped ions where the coherent coupling and the rate of dissipation can be individually controlled and adjusted over a wide range. Thanks to this controllability, our work opens a possibility to investigate potentially rich dynamics associated with a dissipative phase transition.
\end{abstract}

\maketitle
\pagebreak

\section{Introduction} 
Quantum optical systems have emerged as a promising platform to investigate the physics of many-body systems and phase transitions~\cite{Hartmann:2006kv,Greentree:2006jg,Angelakis:2007ho,Hartmann:2016er, Houck:2012iq,Blatt:2012gw,Carusotto:2013gh}. They typically consist of matter represented by two- or few-level systems interacting with quantized light fields or motional degree of freedom, i.e., quantum harmonic oscillators, which in experiments are subject to dissipation. In addition to this intrinsic open nature of these systems, the possibility to bring the systems out of equilibrium in a controlled manner allows one to explore a broad range of nonequilibrium phenomena that has remained difficult to access and yet is vital to advance the understanding of nonequilibrium many-body physics. For example, recent experiments have observed dissipative phase transitions (DPTs) in a BEC trapped in a cavity~\cite{Baumann:2010js,Baumann:2011io,Brennecke:2013dx,Klinder:2015df,Baden:2014gu}, semiconductor microcavity~\cite{Rodriguez:2017by}, and  superconducting circuits~\cite{Fitzpatrick:2017cv,Fink:2017ii}, which are abrupt and nonanalytical changes of the steady state due to the competition among coherent interactions, external drivings, and dissipations.

Another fundamental property of a quantum harmonic oscillator is that its Hilbert space dimension is unbounded. It has been recently pointed out that this can give rise to a sharp notion of phases and phase transitions even in a coupled system of single oscillator and single qubit~\cite{Hwang:2015eq,Hwang:2016cb}. The underlying principle of this so-called finite-component system phase transition is that the ultrastrong qubit-oscillator coupling together with the extremely large detuning achieves a thermodynamic limit of diverging oscillator excitations, in which a nonanalytic change of the ground state may occur. These works, however, have so far been limited to closed systems~\cite{Hwang:2015eq,Hwang:2016cb,Puebla:2016dv,Bakemeier:2012ja,Ashhab:2013ke,Larson:2017kf} despite the intrinsic open nature of harmonic oscillators in experiments. It is therefore an important open question to understand whether it is possible for a finite-component quantum system to reach the thermodynamic limit of diverging excitations through the simultaneously large detuning and coupling even in the presence of dissipation, and, if so, what are the universal properties of a phase transition appearing in such a limit of an open quantum system. 

In this work, we show that a single damped harmonic oscillator coupled to a single qubit, described by an open quantum Rabi model (QRM), undergoes a second-order DPT due to the interplay between the ultrastrong, highly detuned qubit-oscillator coupling and the oscillator damping. In the infinite-$\eta$ limit~\cite{Hwang:2015eq,Hwang:2016cb}, where $\eta$ is the qubit frequency divided by the oscillator frequency, we analytically show the vanishing of the asymptotic decay rate at the critical point, a direct manifestation of the closing gap of the Liouvillian and a hallmark of DPTs~\cite{Kessler:2012hb,Fitzpatrick:2017cv}. This is accompanied by the diverging oscillator population of the steady state at the critical point due to the counter-rotating terms that counteract the loss, even in the absence of external driving fields. Therefore, our study shows that achieving the thermodynamic limit of infinite excitations, in which a finite-component quantum system is allowed to exhibit a nonanalytical change, through the large detuning ($\eta$) and the large coupling strength, is a universal principle working for both closed and open systems.

Moreover, we study the effect of quantum fluctuations due to finite $\eta$ on the DPT, which introduces a nonquadratic interaction for the oscillator to the master equation and makes it no longer amenable to analytical solutions in general. We overcome this challenge by employing the Keldysh path-integral approach~\cite{Kamenev:2009gw,Torre:2013gc} and find analytic expressions for the finite-$\eta$ scaling exponents and reveal the nonequilibrium scaling function, which are identical with those of the open Dicke model. Our analysis demonstrates that the open QRM and the open Dicke model~\cite{Dimer:2007da,Nagy:2010dr,Nagy:2011fu,Oztop:2012bo,Torre:2013gc,Zou:2014cy,Nagy:2015ix,Lang:2016bx,Kirton:2017fx,Scarlatella:2016ug} belong to the same universality class. This finding generalizes the previous studies that the closed QRM and the closed Dicke model belong to the same universality and that the frequency ratio $\eta$ and the number of spins $N$ play the identical role in their respective phase transitions to the setting of a quantum open system. Moreover, the analytical results for finite-$\eta$ scaling relations plays a crucial role in our proposal for observing the DPT of open QRM.

Finally, we propose a method to observe the DPT of the open QRM in a system of two trapped ions. In our scheme, a collective motional mode is coupled to the internal levels of one of the ions to implement the coherent Rabi coupling~\cite{Puebla:2017gq}, while the second ion is used to introduce a controlled amount of damping  via standard laser-cooling techniques~\cite{Home:2013th,Lemmer:2017tj}. The key feature of our scheme is that the damping rate is controllable and can be turned on and off; this opens an exciting possibility for a controlled switch from a quantum phase transition to a DPT and vice versa, a scenario that is not achievable in other setups previously used to realize DPTs in cavity QED systems~\cite{Baumann:2010js,Baumann:2011io,Brennecke:2013dx,Klinder:2015df,Fitzpatrick:2017cv,Fink:2017ii}. Our analysis on the effects of noise demonstrates that the verification of the DPT of the open QRM through the measurement of the finite-$\eta$ scaling exponent is feasible with current technology.

The paper is organized as follows. In Sec.~\ref{openQRM}, we introduce the open QRM and perform a semiclassical analysis of the model which shows an instability of the soft mode and a bifurcation in the steady state, a typical manifestation of a dissipative phase transition at a mean-field level. In Sec.~\ref{DPT}, we present a full quantum mechanical solution for the open QRM in the limit $\eta\rightarrow\infty$ and show that it undergoes a DPT. In this limit, the effective master equation becomes quadratic and therefore we readily find analytical solutions for mean amplitudes, fluctuations as well as the asymptotic decay rate. In Sec.~\ref{Keldysh}, the effects of finite $\eta$ are investigated. We employ the Keldysh approach to predict analytic expressions for the finite-$\eta$ scaling exponent and nonequilibrium scaling function; we also confirm the analytical predictions by numerically solving the master equation.  In Sec.~\ref{universality}, we determine the universality class of the open QRM. In Sec.~\ref{Implementation}, we propose a scheme based on two trapped ions to realize the open QRM with controllable coherent interaction and dissipation and demonstrate the feasibility of observing the DPT of the open QRM in a realistic experimental setup. Finally, we conclude our paper in Sec.~\ref{Conclusion}.

\section{The open quantum Rabi model} 
\label{openQRM}
The model considered in this paper is the open-system version of the QRM, described by a master equation,
\begin{equation}
\label{lindblad_maintext}
\dot\rho=\mathcal{L}[\rho]=-i[\Hr,\rho]+\kappa\mathcal{D}[a],
\end{equation}
where $a$ ($a^\dagger$) and $\kappa$ are the annihilation (creation) operator and the damping rate of a harmonic oscillator, respectively. The dissipator of the oscillator is assumed to be given in Lindblad form, $\mathcal{D}[a]=2a\rho a^\dagger -a^\dagger a \rho-\rho a^\dagger a$, while the coherent dynamics is governed by the Rabi Hamiltonian,
\begin{equation}
\label{rabi}
\Hr=\omega_0 a^\dagger a +\frac{\Omega}{2}\sigma_z -\lambda(a+a^\dagger)\sigma_x,
\end{equation}
where $\sigma_{x,z}$ are Pauli matrices for a two-level system. The oscillator frequency is $\omega_0$, the qubit transition frequency $\Omega$, and $\lambda$ is the coupling strength. It is convenient to introduce a frequency ratio $\eta\equiv\Omega/\omega_0$ and a dimensionless coupling constant $g=2\lambda/\sqrt{\omega_0\Omega}$. The Rabi Hamiltonian can be generalized to many particle models that undergo phase transitions in the thermodynamic limit of infinitely many particles; for example, the Dicke model~\cite{Hepp:1973jt,Emary:2003daa} takes $N$ qubits instead of a single qubit and the Rabi lattice model~\cite{Zheng:2011cb,Schiro:2012hr,Hwang:2013kk,Schiro:2016bx} considers a one-dimensional lattice of coupled oscillators where each oscillator realizes the Rabi Hamiltonian with a local qubit. Interestingly, the Rabi Hamiltonian itself also undergoes a quantum phase transition~\cite{Hwang:2015eq} in the limit of ultrastrong coupling $\lambda/\omega_0\gg1$, and extremely large detuning, $\eta\gg1$, but keeping the coupling constant $g\sim\mathcal{O}(1)$ finite; in the following, we are mainly interested in such a limit.

Note that for an equilibrium system in the ultrastrong coupling regime, the master equation in the form of Eq. (1) is typically not valid because the environment of the oscillator and the qubit cannot be treated independently~\cite{Beaudoin:2011ip,Ridolfo:2012dt,LeBoite:2016fq,LeBoite:2017kd}; however, this can be effectively achieved by a driven trapped ion system, as detailed below, or by cavity-assisted Raman transitions~\cite{Grimsmo:2013id}.  We also emphasize that the effective master equation in Eq.~(\ref{lindblad_maintext}) does not contain any driving terms and the oscillator damping solely competes with the $Z_2$ symmetry preserving qubit-oscillator coupling. This is in stark contrast to first-order DPTs investigated in the driven-dissipative Jaynes-Cummings~\cite{Carmichael:2015cm,Fink:2017ii} or Kerr~\cite{Casteels:2017gp,Rodriguez:2017by} models, where the external driving field used to compensate the oscillator damping explicitly breaks the underlying $U(1)$ symmetry.

Before developing a full quantum mechanical solution of the open QRM, we first perform a semiclassical analysis and find semiclassical steady states of the open QRM. In the limit $\eta\rightarrow\infty$, the semiclassical solution correctly captures the mean-field amplitudes, while neglecting important quantum fluctuations, which will be properly taken into account in the following sections. From the standard Heisenberg-Langevin equations of motion~\cite{Gardiner:2004wr} obtained from the master equation given in Eq.~(\ref{lindblad_maintext}), we neglect quantum fluctuations and factorize expectation values to find a semiclassical equation of motion of the open QRM,
\begin{align}
\label{semieom}
\braket{\dot a}&=-i(\omega_0-i\kappa) \braket{a}-i\lambda( \braket{\sigma_+}+\braket{\sigma_-}),\nonumber\\
\braket{\dot \sigma_+}&=i\Omega\braket{\sigma_+}-i\lambda(\braket{a}+\braket{a}^*)\braket{\sigma_z},\nonumber\\
\braket{\dot \sigma_z}&=-i2\lambda(\braket{a}+\braket{a}^*)(\braket{\sigma_+}-\braket{\sigma_-}).
\end{align}
Therefore, the semiclassical steady state solutions satisfy
\begin{align}
\label{steadyeom}
0&=\left(1-i\frac{\kappa}{\omega_0}\right)\alpha+\frac{g}{2}\left(s_++s_+^*\right),\nonumber\\
0&=-s_++\frac{g}{2}\left(\alpha+\alpha^*\right)s_z,\nonumber\\
0&= g\left(\alpha+\alpha^*\right)\left(s_+-s_+^*\right),
\end{align}
where we have introduced a renormalized steady-state mean amplitude of the oscillator, 
\begin{align}
\alpha\equiv\braket{a}_s/\sqrt{\eta},
\end{align}
and the steady-state qubit expectation values $s_+\equiv\braket{\sigma_+}_s$ and $s_z\equiv\braket{\sigma_z}_s$.

Together with the fact that a pseudoangular momentum is conserved, i.e., $4|s_+|^2+s_z^2=1$, we find that the semiclassical solution of the open QRM exhibits a bifurcation at $g=g_c$, where the critical point $g_c$ is defined as
\begin{align}
g_c=\sqrt{1+\kappa^2/\omega_0^2}.
\end{align}
Below the critical point, $g<g_c$, the only stable solution is a trivial solution with zero mean-field amplitudes,
\begin{align}
\label{trivialsol}
\alpha=0,~s_+=0,
\end{align}
with the qubit being in its ground state $s_z=-1$. Above the critical point $g>g_c$, however, the zero mean-field solution given by Eq.~(\ref{trivialsol}) becomes unstable and bifurcates into two stable solutions with nonzero mean-field solutions,
\begin{align}
\label{meanfield}
\braket{a}=\pm\frac{g\sqrt{\eta}/2}{1-i\frac{\kappa}{\omega_0}}\sqrt{1-\left(g_c/g\right)^4},~s_+=\mp\half\sqrt{1-\left(g_c/g\right)^4}
\end{align}
which accompanies nonzero population of the qubit excited state $s_z=-\frac{g_c^2}{g^2}$. We note that the open QRM preserves the $Z_2$ symmetry, namely, Eq.~(\ref{lindblad_maintext}) is invariant under $\{a\rightarrow-a,\sigma_-\rightarrow-\sigma_-\}$, as there is no explicit driving field that breaks the symmetry. Any symmetry preserving steady-state solution should have zero mean-field amplitude, $\alpha=s_+-=0$. That the non-zero mean-field solutions become stable for $g>g_c$ therefore indicates that a spontaneous symmetry breaking occurs. Moreover, the spontaneous coherence of the oscillator exhibits superradiance in that its amplitude is proportional to $\sqrt{\eta}$ and therefore diverges.

The semiclassical steady-state solution of the open QRM presented here exhibits a bifurcation from a zero mean-field solution to a symmetry-breaking, superradiant mean-field solution, which are reminiscent of the mean-field solution of the open Dicke model~\cite{Dimer:2007da,Oztop:2012bo} that are a manifestation of a dissipative phase transition occurring in the thermodynamic limit of an infinite number of qubits. Our semiclassical analysis here therefore strongly suggests that the open QRM undergoes a DPT and the thermodynamic limit of infinite oscillator excitation is achieved in the limit $\eta\rightarrow\infty$ even in the presence of oscillator damping and in the absence of driving fields that counteract the damping to maintain the finite-density phase. In the next section, we present a full quantum mechanical solution that shows this is indeed the case. It is important to note that in what follows we keep the harmonic oscillator frequency $\omega_0$ finite when we take the limit $\eta\rightarrow\infty$, as it determines the energy scale for quantum fluctuations in such a limit.

\section{Dissipative phase transition} 
\label{DPT}
In this section, we find an analytical and full quantum mechanical solution for the steady state of the open QRM and demonstrate that it undergoes a superradiant dissipative phase transition. To this end, we first derive an effective master equation for the limit $\eta\rightarrow\infty$, which becomes quadratic in the oscillator operator $a$.  From the quadratic effective master equation, we solve linear systems of the equation of motion for both first and second moments of the oscillator. As we will see below, the exact solution for the first moment shows the emergence of the superradiant and broken-symmetry phase, which agrees with the semiclassical solution; the second moment shows the diverging fluctuation around the mean-field solution, which establishes the thermodynamic limit of infinite excitations.

\subsection{Normal phase}
Consider a unitary transformation 
\begin{equation}
U_\textrm{np}=\exp[g\sqrt{\eta^{-1}}/2(a+a^\dagger)(\sigma_+-\sigma_-)],
\end{equation}
which has been shown in Ref.~\cite{Hwang:2015eq} to remove from the Rabi Hamiltonian, given by Eq.~(\ref{rabi}), any coupling terms between the qubit states $\ket{\up}$ and $\ket{\down}$ ($\sigma_z\ket{\up(\down)}=+(-)\ket{\up(\down)}$) up to second order in $g$. We apply the unitary transformation $U_\textrm{np}$ to the master equation (\ref{lindblad_maintext}). Then, the transformed Hamiltonian of the coherent part reads $U_\textrm{np}^\dagger H_\textrm{Rabi} U_\textrm{np} =\omega_0a^\dagger a +\frac{\Omega}{2}\sigma_z+(\omega_0g^2/4)(a+a^\dagger)^2\sigma_z$, while the infinitesimal transformation does not affect the dissipator $\mathcal{D}[a]$ (see the Appendix \ref{appendix A}). Upon a projection to the $\ket{\down}$ subspace of the qubit, we obtain an effective master equation 
\begin{equation}
\label{eq:dpt:01}
	\dot\rho_a=-i[\Hnp,\rho_a]+\kappa\mathcal{D}[a]
\end{equation}
 with 
 \begin{equation}
 \Hnp=\omega_0a^\dagger a -(\omega_0g^2/4)(a+a^\dagger)^2
 \end{equation}
  and $\rho_a\equiv\bra{\down}U_\textrm{np}^\dagger\rho U_\textrm{np}\ket{\down}$. 
 
\begin{figure}[t]
\includegraphics[width=\linewidth]{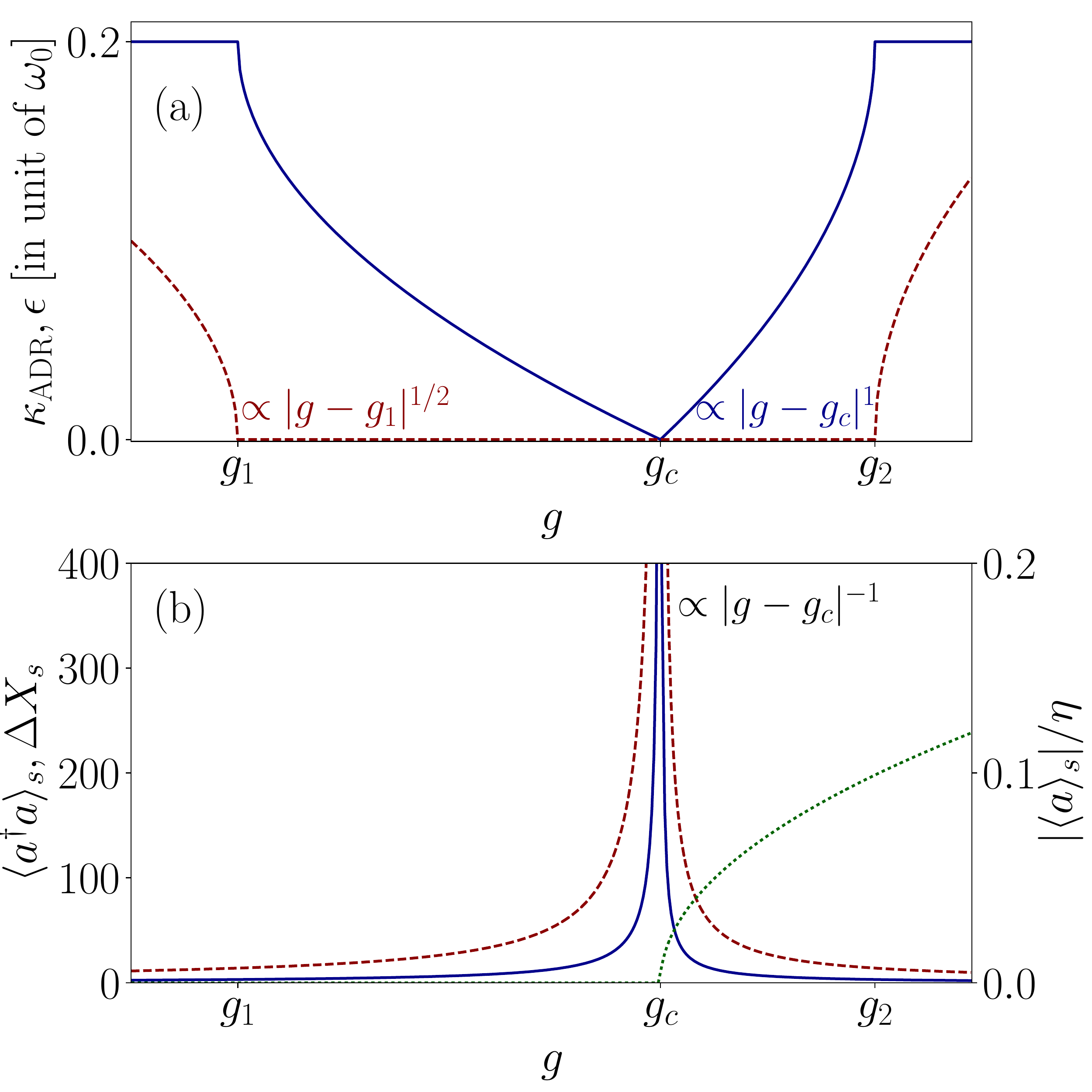}
  \caption{Analytical solutions in the limit $\eta\rightarrow\infty$. (a) The asymptotic decay rate $\kappa_{\rm ADR}$ (solid line) vanishes at $g=g_c$. The excitation energy $\epsilon$ (dashed line) becomes zero for $g_1\leq g\leq g_2$ with $g_1=1$ and $g_2=g_c^{3/2}$, leading to an overdamped dynamics. (b) The steady state expectation values for the order parameter $|\langle a\rangle_s|/\eta$ (dotted line), the oscillator population $\braket{a^\dagger a}_s$ (solid line), and the maximum quadrature variance $\Delta X_s$ (dashed line).  The relevant critical exponents for each quantity are indicated in the figures.}   
  \label{fig:1}
\end{figure}

 From Eq.~(\ref{eq:dpt:01}), we derive a system of linear equations for the mean amplitude $\textbf{u}=(\braket{a},\langle a^\dagger \rangle)^T$,
\begin{align}
\label{eq:dpt:02}
\dot{\textbf{u}}\equiv\textbf{L}_\textrm{np}\textbf{u}=\left(\begin{array}{cc}-i \omega_0(1-\frac{g^2}{2}) -\kappa& i \omega_0\frac{g^2}{2} \\-i \omega_0\frac{g^2}{2} & i \omega_0(1-\frac{g^2}{2})-\kappa\end{array}\right)\textbf{u}.
\end{align}
The eigenvalues of $\textbf{L}_\textrm{np}$ are 
\begin{equation}
\label{eq:dpt:04}
\ell_{\textrm{np},\pm}=-\kappa\pm i\epsilon_\textrm{np},
\end{equation}
where the imaginary part $\epsilon_\textrm{np}=\omega_0\sqrt{1-g^2}$ is the excitation energy in the normal phase of the closed QRM~\cite{Hwang:2015eq}. As long as the real part of $\ell_{\textrm{np},\pm}$ remains negative, the system simply decays to a trivial steady state with zero mean-field amplitudes,
\begin{equation}
\label{eq:dpt:04}
\textbf{u}_{s,\textrm{np}}=(0,0)^T.
\end{equation}
There exists however a critical point $g_c$,
\begin{equation}
g_c=\sqrt{1+\frac{\kappa^2}{\omega_0^2}},
\end{equation}
where the real part of $\ell_{\textrm{np},-}$ becomes zero; see Fig.~\ref{fig:1} (a). For $g>g_c$, $\textrm{Re}[\ell_{\textrm{np},-}]$ becomes positive. This indicates that the trivial solution with zero mean-field amplitude, given by Eq.~(\ref{eq:dpt:04}), is no longer stable and that the mean-field amplitude acquires a nonzero value, thereby breaking the $Z_2$ symmetry of the open QRM.

\subsection{Superradiant phase}
To take into account the emergence of non-zero mean field solutions for $g>g_c$, we first apply the displacement unitary transformation $D[\alpha]=\exp[\alpha a^\dagger -\alpha^* a]$ that displaces the oscillator field, i.e., $a\rightarrow a+\alpha$. A proper choice of $\alpha$ would lead to a stable zero mean-field solution for the steady state in the displaced coordinate. We will see in the following that the semiclassical solution given in Eq.~(\ref{meanfield}) achieves exactly that. With the choice of $\alpha=\pm\alpha_s$ where 
\begin{equation}
\label{eq:dpt:05}
\alpha_s=\frac{g\sqrt{\eta}}{2g_c^2}\left(1+i\frac{\kappa}{\omega_0}\right)\sqrt{1-\left(g_c/g\right)^4},
\end{equation}
we apply the unitary transformation $D[\pm\alpha_s]$ to Eq.~(\ref{lindblad_maintext}) to have
\begin{equation}
\label{eq:dpt:06}
\dot {\bar\rho}_\pm=-i[\bar H_\textrm{Rabi}(\pm\alpha_s),\bar\rho_\pm]+\kappa\mathcal{D}[ a]
\end{equation}
where $\bar\rho_\pm\equiv D^\dagger[\pm\alpha_s] \rho D[\pm\alpha_s]$ and 
\begin{align}
\label{eq:dpt:07}
\bar H_\textrm{Rabi}(\pm\alpha_s)&=\omega_0 a^\dagger a\pm\frac{\omega_0g\sqrt{\eta}}{2}\sqrt{1-\frac{g_c^4}{g^4}}(a+a^\dagger)(1+\tau_z^\pm)\nonumber\\
&-\frac{\omega_0g_c^2\sqrt{\eta}}{2g}(a+a^\dagger)\tau_x^\pm+\frac{\Omega g^2}{2g_c^2}\tau_z^\pm
\end{align}
up to a constant. Here, $\tau_{x,z}^\pm$ are Pauli matrices in a new qubit basis defined as 
\begin{align}
\label{eq:dpt:08}
	\ket{\bar\up_\pm}&=\frac{1}{\sqrt{2}}\left(\sqrt{1+g_c^2/g^2}\ket{\up}\mp\sqrt{1-g_c^2/g^2}\ket{\down}\right),\nonumber\\
	\ket{\bar\down_\pm}&=\frac{1}{\sqrt{2}}\left(\pm\sqrt{1-g_c^2/g^2}\ket{\up}+\sqrt{1+g_c^2/g^2}\ket{\down}\right).
\end{align}
We then find a unitary transformation 
\begin{equation}
\label{eq:dpt:09}
U_\textrm{sp}^\pm=\exp[-\frac{ig_c^4}{2g^3}\sqrt{\eta^{-1}}(a+a^\dagger)\tau_y^\pm]
\end{equation}
to Eq.~(\ref{eq:dpt:06}), which removes any coupling between the new qubit basis states $\ket{\bar\up_\pm}$ and $\ket{\bar\down_\pm}$ from the displaced Hamiltonian given by Eq.~(\ref{eq:dpt:07}) (see the Appendix~\ref{appendix A}). This is followed by a projection onto $\ket{\bar\down_\pm}$ subspace. The resulting effective master equation for the reduced density matrix $\bar\rho_{a,\pm}$ in the superradiant phase reads 
\begin{equation}
\label{eq:dpt:10}
\dot {\bar\rho}_{a,\pm}=-i[H_\textrm{sp},\bar\rho_{a,\pm}]+\kappa\mathcal{D}[ a]
\end{equation}
 with 
 \begin{equation}
 \label{eq:dpt:11}
 H_\textrm{sp}=\omega_0  a^\dagger  a-\frac{\omega_0g_c^6}{4g^4}( a+ a^\dagger)^2.
 \end{equation}
 We provide a detailed derivation of the effective master equation given by Eq.~(\ref{eq:dpt:10}) in Appendix~\ref{appendix A}.
 
From  Eq.~(\ref{eq:dpt:10}), we derive the equation of motion for the mean amplitudes $\textbf{u}=(\braket{a},\langle a^\dagger \rangle)^T$,
\begin{align}
 \label{eq:dpt:12}
\dot{\textbf{u}}=\textbf{L}_\textrm{sp}\textbf{u}.
\end{align}
where
\begin{align}
 \label{eq:dpt:14}
L_\textrm{sp}=\left(\begin{array}{cc}-i \omega_0(1-\frac{g_c^6}{2g^4}) -\kappa& i \omega_0\frac{g_c^6}{2g^4} \\-i \omega_0\frac{g_c^6}{2g^4} & i \omega_0(1-\frac{g_c^6}{2g^4})-\kappa\end{array}\right).
\end{align}
The eigenvalues of $\textbf{L}_\textrm{sp}$ read
\begin{equation}
 \label{eq:dpt:15}
\ell_{\textrm{sp},\pm}=-\kappa\pm i\epsilon_\textrm{sp}
\end{equation}
where $\epsilon_\textrm{sp}=\omega_0\sqrt{1-g_c^6/g^4}$. Note that real values of $\ell_{\textrm{sp},\pm}$ remain negative for $g>g_c$. Therefore, the effective master equation (\ref{eq:dpt:11}) does have a stable zero mean-field amplitude solution when one displaces the oscillator field by $\alpha_s$ determined by the semiclassical solution. From this, we conclude that the open QRM has two possible steady state solutions with spontaneous coherence of the oscillator,
\begin{align}
 \label{eq:dpt:16}
\textbf{u}_{s,\textrm{sp}}=(\pm\alpha_s,\pm\alpha_s^*)^T,
\end{align}
whose amplitude diverges as $\sqrt{\eta}$, leading to a macroscopic occupation of the oscillator population. The steady state solution also spontaneously break the $Z_2$ symmetry of the open QRM.

\subsection{Asymptotic decay rate}
The eigenvalues of systems of the equations of motion in both normal and superradiant phases, given in Eqs.~(\ref{eq:dpt:04}) and (\ref{eq:dpt:15}), show that near the critical point $g_c$, the long-time dynamics is overdamped. More precisely, $\epsilon_\textrm{np}$ for $1<g<g_c$ and $\epsilon_\textrm{sp}$ for $g_c<g<g_c^{3/2}$ become purely imaginary; while this leads to a quantum phase transition at $g=g_c=1$ in the absence of dissipation~\cite{Hwang:2015eq}, here it is balanced with the oscillator damping $\kappa$ and it gives rise to a new time scale, the so-called asymptotic decay rate (ADR)~\cite{Kessler:2012hb}, 
\begin{equation}
\kappa_\textrm{ADR,np}\equiv-\textrm{Re}[\ell_{\textrm{np},-}]=\kappa-\omega_0\sqrt{g^2-1};
\end{equation}
see Fig.~\ref{fig:1} (a). The ADR vanishes at the critical point $g_c=\sqrt{1+\kappa^2/\omega_0^2}$ as
\begin{equation}
\kappa_{\textrm{ADR}}\propto\omega_0|g-g_c|^{\nu_\textrm{ADR}},
\end{equation}
with $\nu_\textrm{ADR}=1$. This is a consequence of the closing of the Liouvillian gap, a hallmark of a DPT~\cite{Kessler:2012hb,Fitzpatrick:2017cv}, at the critical point.

\subsection{Fluctuations}
Now we examine fluctuations of the boson field around the mean amplitude $\textbf{u}_{s,\textrm{np}}$ and $\textbf{u}_{s,\textrm{sp}}$. To this end, we derive systems of linear equations for the boson fluctuations $\textbf{v}=(\langle a^\dagger a \rangle ,\langle a^2 \rangle,\langle a^{\dagger2} \rangle)^T$, which we write as 
\begin{equation}
\dot{\textbf{v}}=\textbf{M}_\textrm{np(sp)} \textbf{v} + \textbf{Y}_\textrm{np(sp)}.
\end{equation}
For the normal phase $g<g_c$, we find
\begin{align}
M_\textrm{np}=i\omega_0\left(\begin{array}{ccc}i\frac{2\kappa}{\omega_0} & -\frac{g^2}{2} & \frac{g^2}{2}\\ g^2 & -2(1-\frac{g^2}{2})+i\frac{2\kappa}{\omega_0}&0 \\ - g^2&0&2(1-\frac{g^2}{2})+i\frac{2\kappa}{\omega_0} \end{array}\right)
\end{align}
and
\begin{align}
Y_\textrm{np}=i\omega_0\left(\begin{array}{c} 0\\\frac{g^2}{2} \\ -\frac{g^2}{2} \end{array}\right).
\end{align}
From this, we derive the steady state solution in the normal phase $\textbf{v}_\textrm{s,np}=-M_\textrm{np}^{-1}Y_\textrm{np}$, which reads
\begin{align}
\label{fluc_ss_np}
\textbf{v}_\textrm{s,np}=\frac{g^2}{8(g_c^2-g^2)}\left(g^2,2-g^2+\frac{2i\kappa}{\omega_0},2-g^2-\frac{2i\kappa}{\omega_0}\right)^T.
\end{align}
For the superradiant phase, we have
\begin{align}
M_\textrm{sp}=i\omega_0\left(\begin{array}{ccc}i\frac{2\kappa}{\omega_0} & -\frac{g_c^6}{2g^4} & \frac{g_c^6}{2g^4}\\ \frac{g_c^6}{g^4} & -2(1-\frac{g_c^6}{2g^4})+i\frac{2\kappa}{\omega_0}&0 \\ - \frac{g_c^6}{g^4}&0&2(1-\frac{g_c^6}{2g^4})+i\frac{2\kappa}{\omega_0} \end{array}\right)
\end{align}
and
\begin{align}
Y_\textrm{sp}=i\omega_0\left(\begin{array}{c} 0\\\frac{g_c^6}{2g^4}\\ -\frac{g_c^6}{2g^4} \end{array}\right),
\end{align}
which leads to the steady state solution
\begin{align}
\label{fluc_ss_sp}
\textbf{v}_\textrm{s,sp}=\frac{g_c^4}{8(g^4-g_c^4)}\left(\frac{g_c^6}{g^4},2-\frac{g_c^6}{g^4}+\frac{2i\kappa}{\omega_0},2-\frac{g_c^6}{g^4}-\frac{2i\kappa}{\omega_0}\right)^T.
\end{align}
Using the analytical solution for the second moments of the oscillator, given in Eqs.~(\ref{fluc_ss_np}) and (\ref{fluc_ss_sp}), we discuss in the following the oscillator population, squeezing, andthe purity of the steady state.
\subsubsection{Oscillator population}
We first consider the oscillator population of the steady state. From the first row of $\textbf{v}_{\textrm{np}(\textrm{sp})}$, we find that it diverges near $g=g_c$ as
\begin{equation}
\label{n_ss}
\langle a^\dagger a\rangle_\textrm{s}\propto|g-g_c|^{-\nu_x},
\end{equation}
with $\nu_x=1$; see Fig.~\ref{fig:1} (b). This so-called photon flux exponent $\nu_x$ of the open QRM differs from $\nu_x=1/2$ of the closed QRM~\cite{Hwang:2015eq}. Note that the presence of a Markovian bath also changes the photon flux exponent of the Dicke model in an identical way, namely, from $\nu_x=1/2$ to $\nu_x=1$~\cite{Torre:2013gc,Oztop:2012bo}. Equation~(\ref{n_ss}) also demonstrates that a thermodynamic limit of diverging oscillator excitations is indeed established in the limit $\eta\rightarrow\infty$ even in the presence of damping and the absence of the explicit driving. This divergence is due to the counter-rotating terms of the Rabi Hamiltonian that counteract the damping and establish a finite-density phase.

\subsubsection{Squeezing}
 Second, we examine the quantum fluctuation along a quadrature variable, $X(\theta)=a e^{-i\theta}+a^\dagger e^{i\theta}$ with $0\leq\theta\leq\pi$. From Eq.~(\ref{fluc_ss_np}), we find the analytical expression for the variance $\Delta X(\theta)=\langle X^2(\theta)\rangle-\langle X(\theta)\rangle^2$ in the normal phase as
 \begin{align}
\label{squeezing}
&\Delta X_{s,\textrm{np}}(\theta)\nonumber\\
&=\frac{g^2}{2(g_c^2-g^2)}\left((1-\frac{g^2}{2})\cos(2\theta)+\frac{\kappa}{\omega_0}\sin(2\theta)+\frac{g^2}{2}\right)+1,
\end{align}
while the expression for $\Delta X_{s,\textrm{sp}}$ can be obtained by simply substituting $g$ from $\Delta X_{s,\textrm{np}}$ with $g_c^3/g^2$. At the critical point, the variance diverges, i.e.,
\begin{align}
\Delta X_{s}(\theta\neq\theta_\textrm{min})\propto|g-g_c|^{-\nu_\Delta},
\end{align}
where $\nu_\Delta=1$ for any $\theta$ [cf. Fig. \ref{fig:1} (b)], except for $\theta^\textrm{np}_\textrm{min}=\pi-\arctan(\frac{\omega_0}{\kappa})$ where we find
\begin{align}
\Delta X_{s}(\theta_\textrm{min},g=g_c)=1/2.
\end{align}
Note that the minimum variance is below the vacuum fluctuation and therefore the steady state exhibits squeezing. However, the product of the maximum and minimum variance, $\Delta X_{s}(\theta_\textrm{max})\Delta X_{s}(\theta_\textrm{min})$ with $\theta_\textrm{max}=\theta_{\textrm{min}}-\pi/2$, diverges at the critical point. This is in stark contrast to the closed QRM where the minimum variance of the ground state at the critical point becomes zero and the maximum variance diverges so that the ground state remains the minimum uncertainty state~\cite{Hwang:2015eq}. 

\subsubsection{Purity}
 Here, we show that the purity of the steady state at the DPT of the open QRM becomes zero. The purity $\mu$ of Gaussian states~\cite{Paris:2003cj}, with our convention of $x=a+a^\dagger$, is given by
\begin{equation}
P=\frac{1}{2\sqrt{\sigma_{xx}\sigma_{pp}-\sigma_{xp}}}
\end{equation}
where 
\begin{align}
\sigma_{xx}&=\half\left(\braket{x^2}-\braket{x}^2\right),\nonumber\\
\sigma_{pp}&=\half\left(\braket{p^2}-\braket{p}^2\right),\nonumber\\
\sigma_{xp}&=\half\left(\half\braket{xp+px}-\braket{x}\braket{p}\right).
\end{align}
From Eqs.~(\ref{fluc_ss_np}) and (\ref{fluc_ss_sp}), we observe that all the second moments, $\braket{a^\dagger a}$, $\braket{a^2}$, and $\braket{a^{\dagger 2}}$ diverge near the critical point with $|g-g_c|^{-1}$. Therefore, it immediately follows that $\sigma_{xx}, \sigma_{pp}, \sigma_{xp}\propto|g-g_c|^{-1}$ and, as a consequence, the purity at the dissipative phase transition vanishes as
\begin{equation}
P(g\sim g_c)\propto|g-g_c|^{\nu_P},
\end{equation}
with $\nu_P=1/2$. Therefore, we conclude that the steady state at the DPT becomes a maximally mixed state.

\section{Keldysh approach for finite-frequency scaling analysis}
\label{Keldysh}
 Having established the DPT of the open QRM in the $\eta\rightarrow\infty$ limit, we now move our focus to the effect of $\eta<\infty$ on the DPT. The results presented here play a very important role for establishing the universality class of the open QRM and for making possible its experimental observation, as we will discuss below. For finite-$\eta$, the quartic interaction, i.e., a term that is propotional to $(a+a^\dagger)^4$, must be taken into account~\cite{Hwang:2015eq} and this makes the master equation no longer amenable to exact analytical solutions. We employ the Keldysh path-integral approach~\cite{Kamenev:2009gw,Torre:2013gc,Sieberer:2016ej} to overcome this challenge and analytically derive the finite-$\eta$ scaling exponents of the open QRM.
 
 \begin{figure}[t]
  \includegraphics[width=\linewidth]{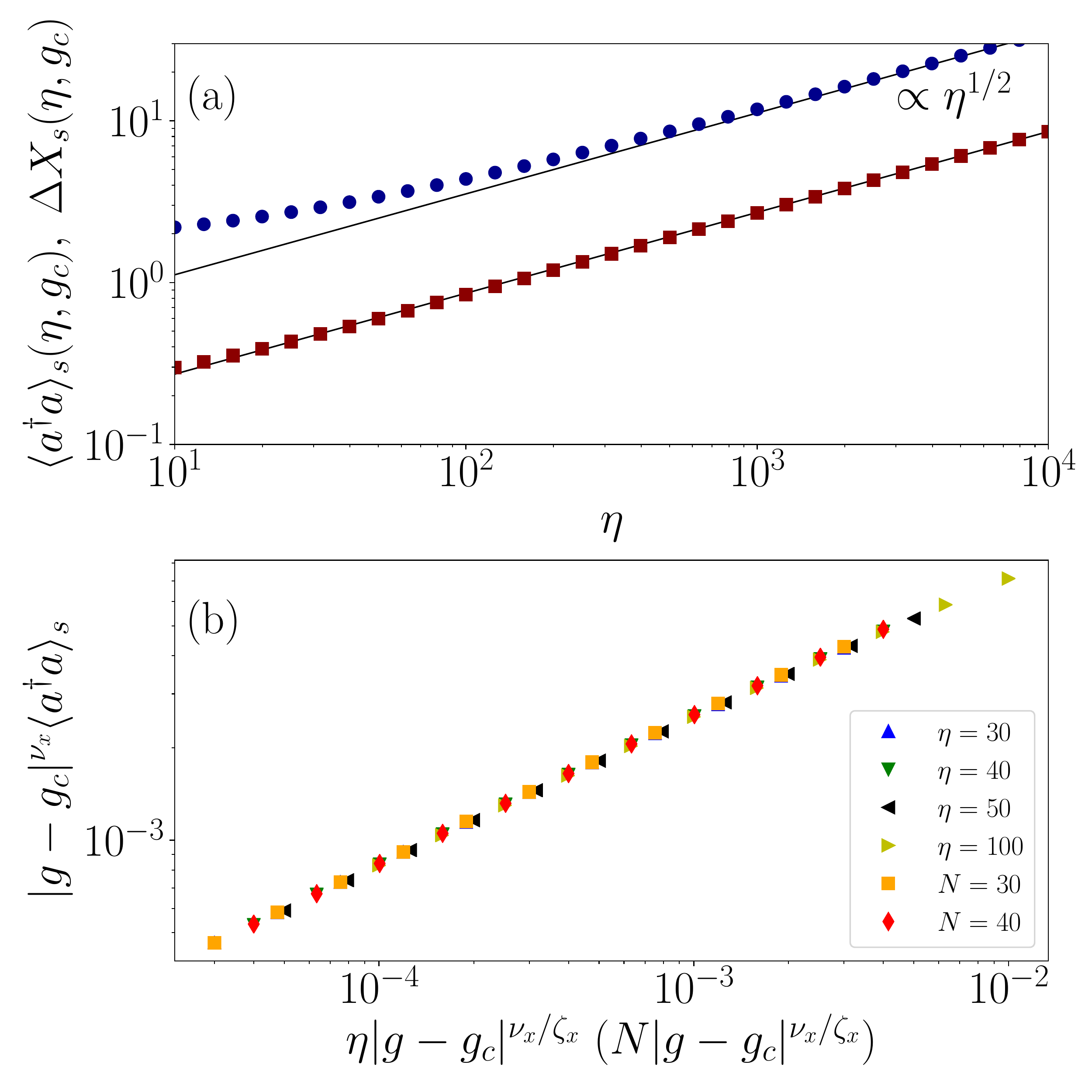} 
  \caption{Finite-$\eta$ scaling relations. (a) Numerical solutions for the oscillator population $\braket{a^\dagger a}_s$ (squares) and the diverging quadrature variance $\Delta X_s$ (circles) of the steady state follows a power-law behavior $\eta^{1/2}$ (solid line), whose exponent is analytically predicted by the Keldysh functional analysis. (b) The rescaled oscillator population $|g-g_c|^{\nu_x}\braket{a^\dagger a}_s$ is plotted as a function of $\eta|g-g_c|^{\nu_x/\zeta_x}$ for the open QRM (triangles) and $N|g-g_c|^{\nu_x/\zeta_x}$ for the open Dicke model (squares). Note that for the open Dicke model, a nonuniversal prefactor $c\sim0.507$ is multiplied to the $y$ axis. All data points collapse onto a single curve.}
   \label{fig:2}
\end{figure}
 
We start from the master equation in the normal phase, given in Eq.~(\ref{eq:dpt:01}),
\begin{align}
\label{eq:keldysh:01}
\dot\rho\equiv\mathcal{L}[\rho]=&-i[\omega_0 a^\dagger a -\frac{\omega_0g^2}{4}(a+a^\dagger)^2,\rho]\nonumber\\&+\kappa(2a \rho a^\dagger - a^\dagger a\rho-\rho a^\dagger a).
\end{align}
Suppose that $\rho(t)$ is a solution to the above equation. The central object in the Keldysh approach is the Keldysh partition function~\cite{Kamenev:2009gw},
\begin{align}
\label{eq:keldysh:02}
Z=\Tr[\rho(t)]=1.
\end{align}
By applying the path integral to the trace of the formal solution, $\rho(t_f)=e^{(t_f-t_i)\mathcal{L}}\rho(t_i)$, and taking a limit of $t_i\rightarrow-\infty$ and $t_f\rightarrow\infty$, we express the partition function $Z$ as
\begin{align}
\label{eq:keldysh:03}
Z=\int d[\alpha_+,\alpha_+^*,\alpha_-,\alpha_-^*]\exp[iS],
\end{align}
where $\alpha_\pm$ are complex numbers defined on the Keldysh contour~\cite{Sieberer:2016ej} and their time dependence is omitted for a compact notation. Here, the action $S$ consists of two parts, namely, 
\begin{align}
S=S_F+S_I.
\end{align}
First, $S_F$ is the free oscillator part with a damping,
\begin{align}
&S_F[\alpha_+,\alpha_+^*,\alpha_-,\alpha_-^*]=\int^\infty_{-\infty}dt\Bigl(\alpha_+^*(i\partial_t-\omega_0)\alpha_+\nonumber\\&-\alpha_-^*(i\partial_t-\omega_0)\alpha_-
-i\kappa[2\alpha_+\alpha_-^*-\alpha_+^*\alpha_+-\alpha_-^*\alpha_-]\Bigr).
\end{align}
Second, $S_I$ is the quadratic interaction part of the oscillator,
\begin{align}
S_I[\alpha_+,\alpha_+^*,\alpha_-,\alpha_-^*]=&\frac{\omega_0g^2}{4}\int^\infty_{-\infty}dt\Bigl((\alpha_+^2+\alpha_+^{*2}+2\alpha_+^*\alpha_++1)\nonumber\\
&-(\alpha_-^2+\alpha_-^{*2}+2\alpha_-^*\alpha_-+1)\Bigr).
\end{align}
After introducing a Keldysh rotation $\acl=(\alpha_++\alpha_-)/\sqrt{2}$ and $\aq=(\alpha_+-\alpha_-)/\sqrt{2}$, we obtain the Keldysh action in the frequency space,
\begin{align}
\label{eq:keldysh:04}
S=\half\int^\infty_{-\infty} \frac{d\omega}{2\pi}V^\dagger(\omega)
\left(
\begin{array}{cc}
0  & [G^\textrm{A}_{2\times2}]^{-1}(\omega)   \\
{ [G^\textrm{R}_{2\times2}]^{-1}(\omega)  } & D_K  \\
\end{array}
\right)
V(\omega).
\end{align}
Here we have introduced the Nambu spinor
\begin{align}
V(\omega)=
\left(
\begin{array}{c}
\acl(\omega)\\
\acl^*(-\omega)\\
\aq(\omega)\\
\aq^*(-\omega)
\end{array}
\right),
\end{align}
the retarded Green's function $G^{R}_{2\times2}$, the advanced Green's function $G^{A}_{2\times2}=(G^{R}_{2\times2})^*$, and the Keldysh Green's function $D_K$. For the normal phase of the open QRM, we find that the retarded Green's function reads
\begin{align}
[G^R_{2\times2}]^{-1}(\omega)=\left(
\begin{array}{cc}
\omega-\omega_0+i\kappa+\Sigma & \Sigma\\
\Sigma &-\omega-\omega_0-i\kappa+\Sigma
\end{array}
\right)
\end{align}
with a self-energy $\Sigma=\omega_0g^2/2$ and the Keldysh Green's function reads
\begin{align}
D_K(\omega)=\left(
\begin{array}{cc}
2i\kappa& 0\\
0 &2i\kappa
\end{array}
\right).
\end{align}

Before analyzing the finite-$\eta$ effect, let us note that the characteristic frequencies of the system are given by $\textrm{det}[G^R_{2\times2}]^{-1}(\omega)=0$, which leads to
\begin{align}
\omega=-i\kappa\pm\omega_0\sqrt{1-g^2}.
\end{align}
The above frequency is closely related to the eigenvalues $\ell_\textrm{np}$ in Eq.~(\ref{eq:dpt:04}), as $\omega=i\ell_\textrm{np}$, and therefore it correctly captures the vanishing of ADR at $g=g_c$. In fact, we note that all of our findings on the DPT of the open QRM in the $\eta\rightarrow\infty$ limit provided in Sec.~\ref{DPT} can also be obtained from the Keldysh action for the open QRM in Eq.~(\ref{eq:keldysh:04}). In Sec.~\ref{DPT}, we have chosen to present our main results in the $\eta\rightarrow\infty$ limit by solving the equations of motion derived from the effective master equation as it is a more accessible approach for a broader audience.  For $\eta<\infty$, however, it is the Keldysh approach presented here that allows us to make an analytical prediction for the open QRM.

 From the Keldysh action given in Eq.~(\ref{eq:keldysh:04}), we derive the finite-$\eta$ scaling exponent using the idea of scale invariance, following the procedure used for the open Dicke model in Ref.~\cite{Torre:2013gc}. To this end, we perform a change of variables using $\alpha_{\textrm{cl}(\textrm{q})}=\sqrt{\omega_0/2}(x_{\textrm{cl}(\textrm{q})}+ip_{\textrm{cl}(\textrm{q})})$ and $\alpha_{\textrm{cl}(\textrm{q})}^*=\sqrt{\omega_0/2}(x_{\textrm{cl}(\textrm{q})}-ip_{\textrm{cl}(\textrm{q})})$, and then integrate out the $p_{\textrm{cl}}$ and $p_{\textrm{q}}$. After a low-frequency expansion, we obtain the Keldysh action in the time domain as
\begin{align}
S=\half\int^\infty_{-\infty} dt(x_\textrm{cl}(t), x_\textrm{q}(t))
\left(
\begin{array}{cc}
0  & -2i\kappa \partial_t   \\
2i\kappa \partial_t  & 2i\kappa \omega_0(1+\kappa^2/\omega_0^2)  \\
\end{array}
\right)
\left(
\begin{array}{c}
x_\textrm{cl}(t)\\ x_\textrm{q}(t)
\end{array}
\right).
\end{align}
It is straightforward to show that the above action is invariant under the scaling transformation,
\begin{align}
\label{scaling1}
t\rightarrow a t,\quad x_\textrm{cl}(t)\rightarrow\sqrt{a}  x_\textrm{cl}(t),\quad x_\textrm{q}(t)\rightarrow\frac{1}{\sqrt{a}} x_\textrm{q}(t).
\end{align}
The lowest-order contributions of the finite-$\eta$ correction to the quadrature effective Hamiltonian $\Hnp$ are quartic interactions~\cite{Hwang:2015eq}. Thus, the expansion of the open QRM up to $\eta^{-1}$ would yield terms such as $\frac{\lambda}{\eta}\int dt \phi_{cl} \phi_{cl} \phi_{cl} \phi_{q}$ or $\frac{\lambda}{\eta}\int dt \phi_{cl} \phi_{q} \phi_{q} \phi_{q}$~\cite{Torre:2013gc}. For these first-order corrections to be invariant under the same scaling transformation, one has to renormalize $\eta$ as
\begin{align}
\label{scaling2}
\eta\rightarrow a^2 \eta.
\end{align}
It follows that $\frac{1}{\sqrt{\eta}}\braket{x_{cl}^2}$ is invariant under the scaling transformation given by Eqs.~(\ref{scaling1}) and (\ref{scaling2}) and therefore the oscillator population and the quadrature variance of the steady state, which is proportional to  $\langle x_\textrm{cl}^2\rangle$, follow a finite-$\eta$ scaling relation, 
\begin{align}
\label{finitescaling}
\langle a^\dagger a\rangle_s(\eta,g=g_c)\propto\eta^{\zeta_x},\quad\Delta X_x(\eta,g=g_c)\propto\eta^{\zeta_\Delta},
\end{align}
with $\zeta_x=\zeta_\Delta=1/2$. We confirm these predictions on the finite-$\eta$ scaling exponents by numerically solving the master equation of the open QRM in Eq.~(\ref{lindblad_maintext}) for $\eta\gg1$ at $g=g_c$, which shows an excellent agreement with Eq.~(\ref{finitescaling}), as shown in Fig. \ref{fig:2}(a).

\section{Universality class}
\label{universality}
So far, we have demonstrated that the open QRM undergoes a DPT in the infinite-$\eta$ limit and exhibits a finite-$\eta$ scaling in the steady state. We have also found analytical expressions for critical exponents characterizing the criticality of the open QRM. First, the ADR, which describes the overdamped dynamics near the critical point due to the closing of the Liouvillian gap, vanishes as $\kappa_\textrm{ADR}\propto|g-g_c|^{\nu_{\textrm{ADR}}}$ with $\nu_{\textrm{ADR}}=1$. Second, the oscillator population of the steady state with respect to the mean amplitude diverges at $g=g_c$ as $\langle a^\dagger a\rangle_s\propto|g-g_c|^{-\nu_x}$ with $\nu_x=1$ for $\eta\rightarrow\infty$ and as $\langle a^\dagger a\rangle_s\propto\eta^{\zeta_x}$ with $\zeta_x=1/2$ for $\eta<\infty$, in contrast to $\nu_x=1/2$ and $\zeta_x=1/3$ for the ground state oscillator population of the closed QRM~\cite{Hwang:2015eq}. All of these critical exponents are identical to the corresponding exponents of the open Dicke model~\cite{Oztop:2012bo}. This observation suggests that the open-system version of the QRM and Dicke model belong to the same universality class. Moreover, the correspondence between the frequency ratio $\eta$ and the number of atoms $N$ in the qubit-oscillator systems demonstrated for a closed system~\cite{Hwang:2015eq,Hwang:2016cb} holds also for an open quantum system~\cite{Hwang:2015eq,Hwang:2016cb}. To determine the universality class of the open QRM and to corroborate that the open QRM and the open Dicke model belongs to the same universality class~\cite{Stanley:1999fd}, we calculate nonequilibrium scaling functions of both the open QRM and the open Dicke model. Together with analytical expressions for the critical exponents for both $\eta\rightarrow\infty$ and $\eta<\infty$, given in Eqs.~(\ref{n_ss}) and Eq.~(\ref{finitescaling}), respectively, we use the scaling hypothesis~\cite{Fisher:1972tv,Botet:1982ju} to find a scaling transformation that reveals the nonequilibrium scaling function for the steady-state oscillator population of the open QRM as
\begin{equation}
\label{scalingft}
	|g-g_c|^{\nu_x}\braket{a^\dagger a}_s(\eta,g)=F_n(\eta|g-g_c|^{\nu_x/\zeta_x}).
\end{equation}
In Fig.~\ref{fig:2} (b), we numerically calculate the steady state expectation value $\langle a^\dagger a \rangle_s (\eta,g)$ from Eq.~(\ref{lindblad_maintext}) for different values of $\eta$ and $g$ satisfying $\eta\gg1$ and $g\sim g_c$ and then plot the rescaled oscillator population $|g-g_c|^{\nu_x}\langle a^\dagger a \rangle_s$ as a function of a rescaled coupling strength $\eta|g-g_c|^{\nu_x/\zeta_x}$. The single curve on which all the data points collapse is the nonequilibrium scaling function. 

We perform the same scaling transformation with Eq.~(\ref{scalingft}) for the open Dicke model where $\eta$ is replaced by $N$, i.e., 
\begin{equation}
|g-g_c|^{\nu_x}\braket{a^\dagger a}_s(N,g)=cF_{n}^\textrm{Dicke}(N|g-g_c|^{\nu_x/\zeta_x}). 
\end{equation}
The form of the scaling transformation above agrees with the one presented in Ref.~\cite{Konya:2012gf}, in which the value of $\nu_x/\zeta_x$ has been obtained through a numerical calculation that deviates slightly from the analytical value $\nu_x/\zeta_x=2$ used here. As shown in Fig.~\ref{fig:2} (b), $F_n$ and $F_{n}^\textrm{Dicke}$ are identical, and thus universal, and the calculated nonuniversal prefactor is $c\sim0.507$. This confirms that the open QRM and the open Dicke model belongs to the same universality class. Note that the ratio of critical exponent $\xi=\nu_x/\zeta_x$ appearing in the argument of scaling functions $F_n$ and $F_n^\textrm{Dicke}$ is sometimes referred to as a coherence number~\cite{Botet:1982ju} for the models without spatial degrees of freedom or for the infinitely-coordinated systems. While $\nu_x$ and $\zeta_x$ are specific to observables, which in this case is the oscillator population, the coherence number $\xi$ is specific to the model and is observable independent. For the open QRM and the open Dicke model, we find $\xi=2$. For the closed QRM and closed Dicke model, on the other hand, we have $\xi=3/2$~\cite{Hwang:2015eq,Emary:2003daa,Vidal:2007ex}. Finally, while we have focused on the oscillator population, the same scaling analysis can be applied to other observables such as $\Delta X_x$ which would lead to the identical scaling function for both models.

\begin{figure}[t]
  \includegraphics[width=\linewidth]{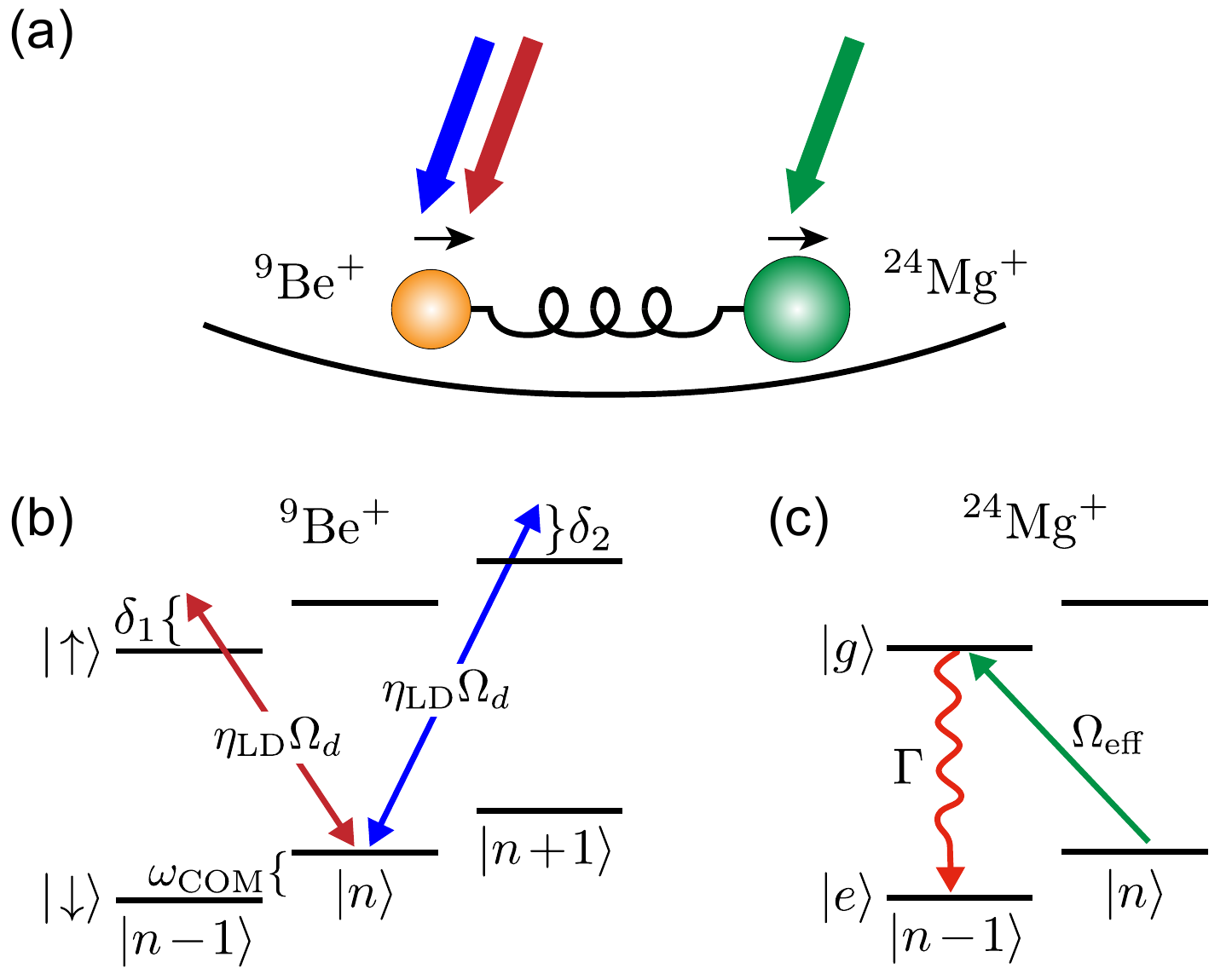} 
  \caption{(a) Realization of the open QRM using a $^{9}{\rm Be^+-^{24}}{\rm Mg}^+$ ion pair in a linear trap. (b) Two lasers are applied to the $^{9}{\rm Be^+}$ in order to drive the blue and red sideband transitions with detuning $\delta_1$ and $\delta_2$, respectively, and a Rabi frequency $\eta_{\rm LD}\Omega_d$. This creates the coherent Rabi coupling between two hyperfine states of $^{9}{\rm Be^+}$ and the center-of-mass (COM) mode. (c) The $^{24}{\rm Mg}^+$ ion is used to implement a tunable phonon damping rate $\kappa=2\Omega_e^2/\Gamma$ via a weak red-sideband excitation to an optically excited state $|e\rangle$. 
}
   \label{fig:3}
\end{figure}

\section{Implementation based on two trapped-ions}
\label{Implementation}
We propose a method for an experimental observation of the predicted DPT in the open QRM using two trapped ions in a linear trap. See Fig.~\ref{fig:3} for a schematic of our proposal. While the proposed scheme is not specific to a certain species of ions, to closely examine the feasibility we consider a specific setup with a mixed species ion pair $^{9}{\rm Be^+-^{24}}{\rm Mg}^+$~\cite{Lin:2013be,Tan:2015ka}. We choose the common center of mass mode as the oscillator of the QRM. All other vibration modes are far separated in frequency and can be neglected. The hyperfine states of  $^{9}{\rm Be^+}$, $\ket{F=2,m_F=0}$ and $\ket{F=1,m_F=1}$, form a qubit, which can be coupled to the motional mode using coherent stimulated Raman transitions~\cite{Monroe:1995ia}. After moving to the interaction picture with respect to the bare qubit and oscillator dynamics, followed by a rotating wave approximation (RWA), the interaction Hamiltonian between the oscillator and qubit in the Lamb-Dicke limit is $H_I=\eta_{\rm LD}\Omega_d\sigma^+(a e^{i\delta_1 t}+a^\dagger e^{i\delta_2 t})+h.c.$, where we have considered two lasers driving both the blue- and red-sideband transition and $\delta_1$ ($\delta_2$) is a detuning of the driving laser with respect to the red-(blue)-sideband transition, $\Omega_d$ is the Rabi frequency, and $\eta_{\rm LD}\sim0.15$ is the Lamb-Dicke parameter~\cite{Lin:2013be,Tan:2015ka}. In the rotating frame, where $H_I$ becomes time independent, $H_I$ takes the form of $H_{\rm Rabi}$ with $\omega_0=(\delta_2-\delta_1)/2$, $\Omega=(\delta_1+\delta_2)/2$, and $\lambda=\eta_{\rm LD}\Omega_d$~\cite{Pedernales:2015cj,Puebla:2017gq}. 

To the above scheme, which allows one to observe the quantum phase transition (QPT) of the closed QRM~\cite{Puebla:2017gq}, one can \emph{controllably} introduce a dissipation to the oscillator, thereby switching the system from probing the QPT of the closed QRM to the DPT of the open QRM. We propose to achieve this by laser-cooling the motional mode with the help of the second $^{24}{\rm Mg}^+$ ion. The sympathetic cooling of the in-phase mode using $^{24}{\rm Mg}^+$ has already been experimentally achieved~\cite{Lin:2013be,Tan:2015ka}. In this setting, the cooling of the normal modes introduces the oscillator damping~\cite{Lemmer:2017tj} and the $^{9}{\rm Be^+-^{24}}{\rm Mg}^+$ ion pair now realizes the dynamics described by Eq.~(\ref{lindblad_maintext}), i.e., the open QRM. The finite-$\eta$ scaling of the phonon number in the steady state is a quantity to be measured and it already emerges for $50\lesssim\eta\lesssim100$ [cf. Fig~\ref{fig:2} (a)]; a possible set of parameters to realize this range of $\eta$ is $\omega_0/2\pi=500 {\rm Hz}$ and $25{\rm kHz}\leq\Omega/2\pi\leq50{\rm kHz}$. The sympathetic cooling rates as high as tens of kHz have been achieved~\cite{Lin:2013be} and here we set the cooling rate $2\kappa/2\pi=200{\rm Hz}$ so that we have $\kappa/\omega_0=0.2$ for the parameters used here, as assumed throughout the paper. The critical coupling strength $\lambda_c=0.5\omega_0\sqrt{\eta}\sqrt{1+(\kappa/\omega_0)^2}$ is then realized in a range of $1.8{\rm kHz}<\eta_{\rm LD}\Omega_d/2\pi<2.5{\rm kHz}$. All of these parameters are within the range of validity of the RWA and Lamb-Dicke limit~\cite{Puebla:2017gq}.

\begin{figure}[t]
\includegraphics[width=\linewidth]{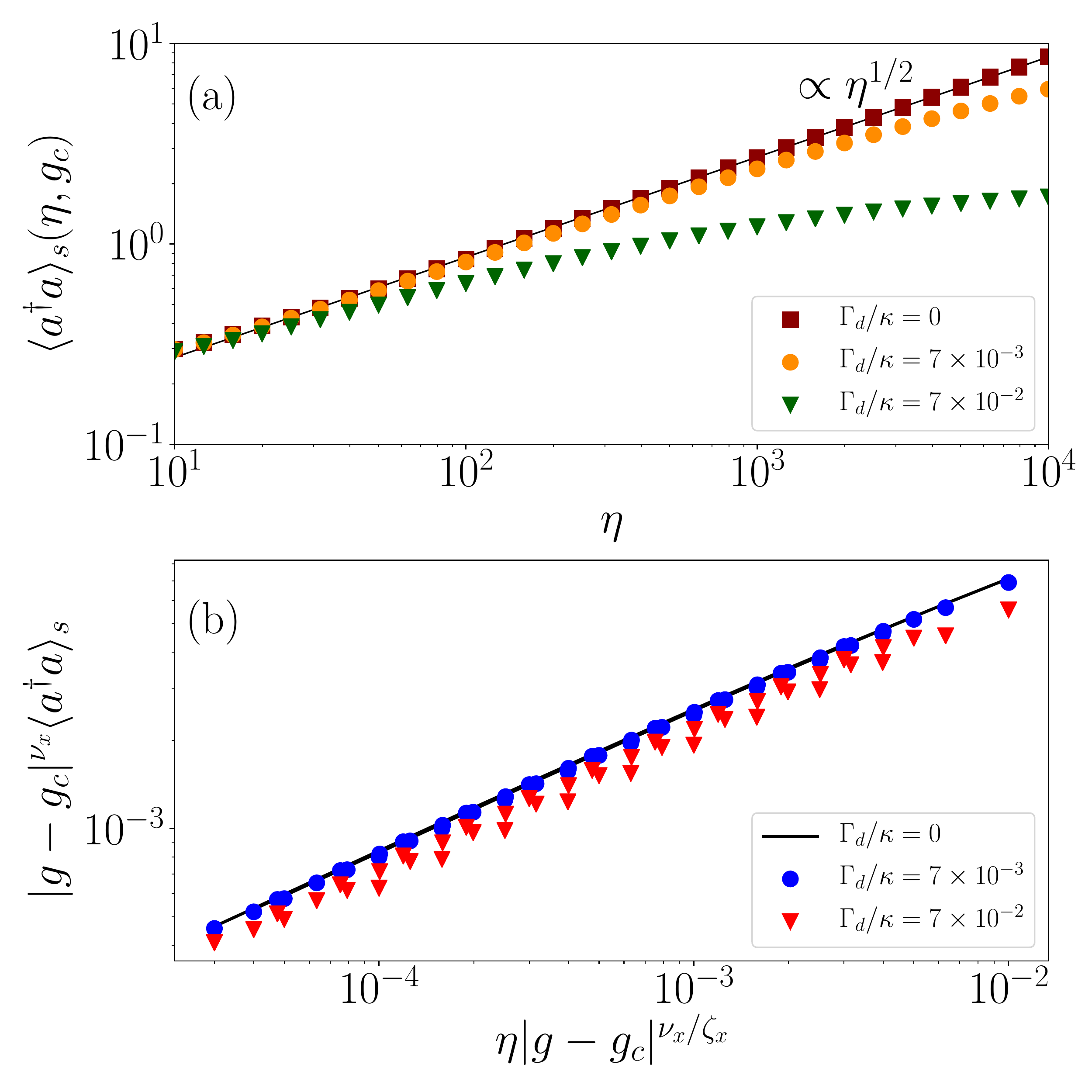} 
  \caption{The effect of dephasing noise on finite-$\eta$ scaling relations. (a) Numerical solutions for the oscillator population $\braket{a^\dagger a}_s$ of the steady state for different values of dephasing rate, $\Gamma_d/\kappa=0$ (squares), $\Gamma_d/\kappa=7\times10^{-3}$ (circle), and $7\times10^{-2}$ (triangles). (b) The rescaled steady state oscillator population $|g-g_c|^{\nu_x}\braket{a^\dagger a}_s$ is plotted as a function of $\eta|g-g_c|^{\nu_x/\zeta_x}$. The solid line is the non-equilibrium scaling function of the open QRM without any dephasing nose. While the non-equilbrium scaling function is still intact with the dephasing rate $\Gamma_d/\kappa=7\times10^{-3}$ (circle), the data no longer collapses on to a single curve for  $\Gamma_d/\kappa=7\times10^{-2}$ (triangle). For all data, the same set of values for $\eta$ and $g$ is used. }   
  \label{fig:S1}
\end{figure}

 Finally, we examine the effect of dephasing noise of the qubit on the DPT. The master equation including the qubit dephasing noise reads
\begin{equation}
\label{lindblad}
\dot\rho=\mathcal{L}[\rho]=-i[\Hr,\rho]+\kappa\mathcal{D}[a]+\Gamma_\textrm{d}\mathcal{D}[\sigma_z]
\end{equation}
where $\mathcal{D}[x]=2x\rho x^\dagger -x^\dagger x \rho-\rho x^\dagger x$. For numerical simulations, we choose $\Gamma_d/\kappa=7\times10^{-3}$ and $\Gamma_d/\kappa=7\times10^{-2}$. The former corresponds to the dephasing rate reported in Ref.~\cite{Tan:2015ka} for an experimental setup based on the hyperfine states of $^{9}{\rm Be^+}$. From Figs.~\ref{fig:S1} (a) and (b), we conclude that for this experimentally accessible dephasing rate of $\Gamma_d/\kappa=7\times10^{-3}$, one can quantitatively measure both the finite-$\eta$ scaling exponent for the oscillator population $\zeta_x=1/2$ and the universal non-equilibrium scaling function for the experimentally accessible values of the frequency ratio $50\lesssim\eta\lesssim100$. For a stronger dephasing rate, e.g., $\Gamma_d/\kappa=7\times10^{-2}$, the scaling relations are strongly modified by the dephasing noise. Therefore, choosing an ion with a long coherence time, in this case the hyperfine states of $^{9}{\rm Be^+}$, is highly advantageous in this regard.

We emphasize that the oscillator damping in our proposal is highly tunable; therefore, one could realize either the QPT of the closed QRM or the DPT of the open QRM in the same experimental setup and even switch from one another suddenly or adiabatically in time. This remarkable controllability of the dissipation in an experimental realization of a DPT is not available in any currently available cavity QED system with optical and microwave photons~\cite{Baumann:2010js,Baumann:2011io,Brennecke:2013dx,Klinder:2015df,Baden:2014gu,Fitzpatrick:2017cv, Fink:2017ii, Rodriguez:2017by}. It opens an exciting opportunity to experimentally investigate the dynamics of DPT and to examine the crossover between a QPT and a DPT. 

\section{Conclusion} 
\label{Conclusion}

In conclusion, we have demonstrated that the open QRM undergoes a DPT, established its universality class, and proposed an experimental scheme based on ion-traps where the predicted DPT can be induced by a motional cooling of ions. Our work shows that the notion of phase transitions in a finite-component system of a coupled oscillator and spin extends to an open quantum system and provides a theoretical and experimental framework to systematically investigate the nature of dissipative phase transitions and its dynamics in a small, fully controlled open quantum system. The gained understanding in the proposed setting may have a far-reaching implication for a wide range of experimental setups~\cite{Baumann:2010js,Baumann:2011io,Brennecke:2013dx,Klinder:2015df,Baden:2014gu} thanks to the universality established here.

\begin{acknowledgments}
This work was supported by the ERC
synergy grant BioQ,  the EU project QUCHIP, and the COST Action NQO (MP1403). P.R. acknowledges support from 
the Austrian Science Fund (FWF) through SFB FOQUS
F40 and the START grant Y 591-N16. M.-J.H acknowledges discussions with A. Lemmer and M. S. Kim. The numerical calculation is performed using QuTip~\cite{Johansson:2013gb}.
\end{acknowledgments}

\appendix

\section{Derivation of effective master equations}
\label{appendix A}
In this section, we derive the effective master equation of the open quantum Rabi model (QRM)  in the $\eta\rightarrow\infty$ limit for both the normal and superradiant phase. 

First, for the normal phase, we consider a unitary transformation $U_\textrm{np}=\exp[g\sqrt{\eta^{-1}}/2(a+a^\dagger)(\sigma_+-\sigma_-)]$ that has been used to derive the effective Hamiltonian of the closed QRM in Ref.~\cite{Hwang:2015eq} and apply to the master equation of the open QRM, i.e.,
\begin{equation}
U_\textrm{np}^\dagger \dot\rho U_\textrm{np}=-iU_\textrm{np}^\dagger [\Hr,\rho] U_\textrm{np}+\kappa U_\textrm{np}^\dagger \mathcal{D}[a]U_\textrm{np}.
\end{equation}
In the $\eta\rightarrow\infty$ limit, the coherent part in the above equation becomes
\begin{align}
&-iU_\textrm{np}^\dagger [\Hr,\rho] U_\textrm{np}\nonumber\\
&=-i[\omega_0a^\dagger a +\frac{\Omega}{2}\sigma_z+\frac{\omega_0g^2}{4}(a+a^\dagger)^2\sigma_z,U_\textrm{np}^\dagger \rho U_\textrm{np}],
\end{align}
which follows from $U_\textrm{np}^\dagger \Hr U_\textrm{np}=\omega_0a^\dagger a +\frac{\Omega}{2}\sigma_z+(\omega_0g^2/4)(a+a^\dagger)^2\sigma_z+\mathcal{O}(\eta^{-1/2})$~\cite{Hwang:2015eq}. In the zeroth order in $\eta$, the dissipator part does not change and all the corrections have an order higher than $\eta^{-1/2}$, which becomes zero in the considered limit. Therefore, the transformed master equation is diagonal in the spin basis $\ket{\up}$ and $\ket{\down}$, and upon the projection onto the spin $\ket{\down}$ subspace we obtain the effective master equation,
\begin{equation}
\label{master_np}
\dot\rho_a=-i[\omega_0a^\dagger a -\frac{\omega_0g^2}{4}(a+a^\dagger)^2,\rho_a] +\kappa\mathcal{D}[a]\rho_a.
\end{equation}
where $\rho_a\equiv\bra{\down}U_\textrm{np}^\dagger\rho U_\textrm{np}\ket{\down}$. 

Second, we now derive the effective master equation for the superradiant phase. We begin by applying the displacement unitary transformation to Eq.~(\ref{lindblad_maintext}) with 
$D[\alpha]=\exp[\alpha a^\dagger -\alpha^* a]$, which leads to
\begin{align}
\dot {\bar\rho}=&-i[D^\dagger[\alpha]  H_\textrm{Rabi}D[\alpha]+i\kappa(\alpha^*a-\alpha a^\dagger),\bar\rho]\nonumber\\&+\kappa(2a \bar\rho a^\dagger - a^\dagger a\bar\rho-\bar\rho a^\dagger a)
\end{align}
where $\bar\rho\equiv D^\dagger[\alpha] \rho D[\alpha]$.
Upon choosing $\alpha=\pm\alpha_s$ where $\alpha_s$ is a mean-field amplitude of the field of the steady state given in Eq.~(\ref{meanfield}),
the master equation becomes
\begin{equation}
\dot {\bar\rho}_\pm=-i[\bar H_\textrm{Rabi}(\pm\alpha_s),\bar\rho_\pm]+\kappa(2a \bar\rho_\pm a^\dagger - a^\dagger a\bar\rho_\pm-\bar\rho_\pm a^\dagger a)
\end{equation}
where the coherent part reads
\begin{align}
\label{displaced}
\bar H_\textrm{Rabi}(\pm\alpha_s)&=\omega_0 a^\dagger a+\omega_0|\alpha_s|^2\nonumber\\&\pm\frac{g\sqrt{\omega_0\Omega}}{2}\sqrt{1-\frac{g_c^4}{g^4}}(a+a^\dagger) -\lambda(a+a^\dagger)\sigma_x\nonumber\\&\mp2\lambda \textrm{Re}[\alpha_s]\sigma_x+\frac{\Omega}{2}\sigma_z.
\end{align}
The spin part of $\bar H_\textrm{Rabi}(\pm\alpha_s)$, i.e., the last two terms of the above equation, becomes diagonal in the following new spin basis,
\begin{align}
\ket{\bar\up_\pm}&=\frac{1}{\sqrt{2}}\left(\sqrt{1+\frac{g_c^2}{g^2}}\ket{\up}\mp\sqrt{1-\frac{g_c^2}{g^2}}\ket{\down}\right),\nonumber\\
\ket{\bar\down_\pm}&=\frac{1}{\sqrt{2}}\left(\pm\sqrt{1-\frac{g_c^2}{g^2}}\ket{\up}+\sqrt{1+\frac{g_c^2}{g^2}}\ket{\down}\right).
\end{align}
Let us define Pauli matrices in the new spin basis $\tau_x^\pm=\ket{\bar\up_\pm}\bra{\bar\down_\pm}+\ket{\bar\down_\pm}\bra{\bar\up_\pm}$, $\tau_y^\pm=i(\ket{\bar\up_\pm}\bra{\bar\down_\pm}-\ket{\bar\down_\pm}\bra{\bar\up_\pm})$ and $\tau_z^\pm=\ket{\bar\up_\pm}\bra{\bar\up_\pm}-\ket{\bar\down_\pm}\bra{\bar\down_\pm}$, which are related to the Pauli matrices in the original spin basis in the following way,
\begin{align}
    \sigma_+&=\pm\half\left(-\sqrt{1-\frac{g_c^4}{g^{4}}}\tau_z^\pm\pm \frac{g_c^2}{g^{2}}\tau_x^\pm\pm i\tau_y^\pm\right),\nonumber\\
    \sigma_x&=\pm\left(-\sqrt{1-\frac{g_c^4}{g^{4}}}\tau_z^\pm\pm \frac{g_c^2}{g^{2}}\tau_x^\pm\right).
\end{align}
The displaced Hamiltonian in Eq.~(\ref{displaced}) in this new spin basis reads
\begin{align}
\bar H_\textrm{Rabi}(\pm\alpha_s)\equiv\bar H_0^\pm-\bar V^\pm,
\end{align}
where
\begin{align}
\bar H_0^\pm
=&\omega_0 a^\dagger a+\omega_0|\alpha_s|^2\pm\frac{g\sqrt{\omega_0\Omega}}{2}\sqrt{1-\frac{g_c^4}{g^4}},(a+a^\dagger)(1+\tau_z^\pm)\nonumber\\&+\frac{\Omega g^2}{2g_c^2}\tau_z^\pm,
\end{align}
and
\begin{align}
\bar V^\pm&=\frac{g_c^2\sqrt{\omega_0\Omega}}{2g}(a+a^\dagger)\tau_x^\pm.
\end{align}
We now find a unitary transformation $U_\textrm{sp}^\pm=e^{-S_\textrm{sp}^\pm}$ which decouples the spin and the oscillator up to the second order in $\bar V^\pm$ following the approach of Ref.~\cite{Hwang:2015eq}. To this end, the generator should satisfy
\begin{align}
[\bar H_0^\pm,S_\textrm{sp}^\pm]=\bar V^\pm,
\end{align}
from which we find
\begin{align}
S_\textrm{sp}^\pm=i\frac{g_c^4}{2g^3}\eta^{-1/2}(a+a^\dagger)\tau_y^\pm+\mathcal{O}\left(\eta^{-1}\right).
\end{align}
Upon this choice of the generator, the transformed Hamiltonian becomes
\begin{align}
\bar H_\textrm{Rabi}(\pm\alpha_s)&=\bar H_0^\pm-\half[\bar V^\pm,S_\textrm{sp}^\pm]+...\nonumber\\
&=\omega_0 a^\dagger a+\omega_0|\alpha_s|^2\nonumber\\
&\quad\pm\frac{g\sqrt{\omega_0\Omega}}{2}\sqrt{1-\frac{g_c^4}{g^4}}(a+a^\dagger)(1+\tau_z^\pm)\nonumber\\
&\quad+\frac{\Omega g^2}{2g_c^2}\tau_z^\pm+\frac{\omega_0g_c^6}{4g^4}(a+a^\dagger)^2\tau_z^\pm+\mathcal{O}\left(\eta^{-\half}\right).
\end{align}
By projecting to the spin subspace of $\ket{\down_\pm}$, we arrive at 
\begin{align}
H_\textrm{sp}&\equiv\bra{\down_\pm}\bar H_\textrm{Rabi}(\pm\alpha_s)\ket{\down_\pm}\nonumber\\&=\omega_0 a^\dagger a-\frac{\omega_0g_c^6}{4g^4}(a+a^\dagger)^2+E_{G,\textrm{sp}}(g),
\end{align}
where the constant energy shift is given by
\begin{align}
E_{G,\textrm{sp}}(g)=\omega_0|\alpha_s|^2-\frac{\Omega g^2}{2g_c^2}=-\frac{\Omega}{4}\left(\frac{g^2}{g_c^2}+\frac{g_c^2}{g^2}\right).
\end{align}
Finally, the effective master equation in the superradiant phase therefore reads
\begin{equation}
\label{master_sp}
\dot {\bar\rho}_\pm=-i[H_\textrm{sp},\bar\rho_\pm]+\kappa(2a \bar\rho_\pm a^\dagger - a^\dagger a\bar\rho_\pm-\bar\rho_\pm a^\dagger a).
\end{equation}
Note that both signs of $\alpha=\pm\alpha_s$ lead to the identical effective Hamiltonian $H_\textrm{sp}$, and thus the identical effective master equations and the doubly degenerate steady states.

\bibliographystyle{apsrev4-1}
\nocite{apsrev41Control}

\bibliography{paper}

\end{document}